\documentclass[prl,showpacs,twocolumn,aps]{revtex4}

\usepackage{amsmath}
\usepackage{amssymb}
\usepackage{latexsym}
\usepackage{amsfonts}
\usepackage{epsfig}
\usepackage{psfrag}
\usepackage{graphicx}
\usepackage{subfigure}

\usepackage{color}

\definecolor{black}{rgb}{0,0,0}
\definecolor{blue}{rgb}{0,0,1}
\definecolor{green}{rgb}{0,1,0}
\definecolor{red}{rgb}{1,0,0}

\newcommand{\ket}[1]{\left|#1\right>}
\newcommand{\bra}[1]{\left<#1\right|}

\newcommand{\f}[1]{\mbox{\boldmath$#1$}}

\newcommand{\bea}{\begin{eqnarray}}
\newcommand{\ea}{\end{eqnarray}}
\newcommand{\eea}{\end{eqnarray}}
\newcommand{\ord}{\,{\cal O}}

\begin{document}

\title{Entangling photons via the double quantum Zeno effect}

\author{Nicolai ten Brinke, Andreas Osterloh, and Ralf Sch\"utzhold$^*$}

\affiliation{Fakult\"at f\"ur Physik, Universit\"at Duisburg-Essen, 
Lotharstrasse 1, 47057 Duisburg, Germany}

\date{\today}

\begin{abstract}
We propose a scheme for entangling two photons via the quantum Zeno effect, 
which describes the inhibition of quantum evolution by frequent measurements 
and is based on the difference between summing amplitudes and probabilities. 
For a given error probability $P_{\rm error}$, our scheme requires that 
the one-photon loss rate $\xi_{1\gamma}$ and the two-photon absorption rate 
$\xi_{2\gamma}$ in some medium satisfy 
$\xi_{1\gamma}/\xi_{2\gamma}=2P_{\rm error}^2/\pi^2$, which is significantly 
improved in comparison to previous approaches.
Again based on the quantum Zeno effect, as well as coherent excitations, 
we present a possibility to fulfill this requirement in an otherwise 
linear optics set-up. 
\end{abstract}


\maketitle

%
%
%
It is well known that photons possess many properties which make them very 
suitable candidates for quantum information processing.
They can be well controlled and manipulated, as well as created and 
measured \cite{DiVincenzo:2000fk}, even down to the single-photon level. 
Furthermore, photons can propagate over relatively long distances without 
significantly coupling to the environment.
However, the latter advantage is also related to their main drawback -- 
it is very hard to make two photons interact. 
Typically, before two photons interact by means of some non-linear 
medium, for example, at least one of them is absorbed. 
This motivates the idea to turn the problem around and to actually exploit 
the absorption in order to make photons interact. 
An indirect way of doing this is realized by the Knill-Laflamme-Milburn
(KLM) gate \cite{Knill:2001vn} which induces interactions (probabilistically)
via entangled ancilla photons and photo-detectors.  
This proposal has been studied in great detail theoretically 
(e.g., \cite{KLM_IMP})
and experimentally (e.g., 
\cite{KLM_EXP}).

A more direct way of exploiting the absorption of photons is to employ the 
quantum Zeno effect which describes the slow-down or even inhibition of 
quantum evolution by repeated measurements \cite{QZE}. 
Imagine a quantum particle in a double-well potential initially confined
to the right well, for example.
After some time $T$, it would tunnel to the left well (and then  back, etc). 
However, if we measure the position of the particle frequently, i.e., after 
very short time intervals $\Delta t\ll T$, it does not tunnel since each 
measurement projects the quantum state back to the right well. 
This is the basic picture of the quantum Zeno effect.
Since the absorption of a photon in some medium is equivalent to measuring
its position, a strong enough absorption probability can actually 
prevent the photon from tunneling/propagating into this medium. 
In contrast to earlier proposals for photon gates based on the quantum Zeno 
effect \cite{FRANSON_AND_LEUNG, MYERS_AND_LEUNG,Huang:2008xq}, 
our set-up is modified and so yields a significantly reduced error 
probability; see Eq.~(\ref{constraint}). 
Furthermore, we propose a realization based on free-space propagation, 
i.e., without waveguides, resonators, or optical fibers etc., which may 
induce losses and decoherence.

\begin{figure}[h]
\begin{center}
%
%
%
\includegraphics[width=1.0\columnwidth]{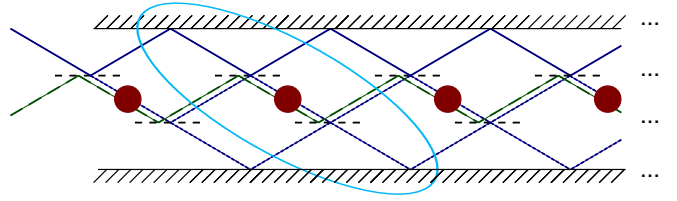}
\caption{Sketch of the macro-structure of the proposed set-up.
Horizontal solid lines (top and bottom) denote perfect mirrors and 
horizontal dashed lines indicate beam splitters.
The target photon (slanting dark blue line) enters the upper branch 
(initial state $\ket{0}$). 
The control photon (slanting dark green dashed line) enters the middle 
branch (if present, i.e, if in the state $\ket{1}$). 
Then both, the control photon and the part of the target photon which 
tunneled through the beam splitter into the middle branch pass the 
two-photon absorbing medium (brown circle).   
If the target photon is not absorbed, it may continue to tunnel to 
the lower branch through the second beam splitter. 
Each segment (tilted light blue oval) consists of an absorbing medium 
and two beam splitters.}
\label{fig:3branch}
\end{center}
\end{figure}

The macro-structure of our set-up is sketched in Figure~\ref{fig:3branch}.
It consists of perfect mirrors and beam splitters with reflectivity 
$\cos\epsilon$ and transmittivity $\sin\epsilon$ where $\epsilon\ll1$.
The target photon enters either the upper branch $\ket{0}$ or the 
lower branch $\ket{1}$. 
The control photon, on the other hand, enters the middle branch 
if it is in the state $\ket{1}$.
The polarization and/or frequency of the control photon is chosen such 
that it is perfectly reflected by the beam splitters and thus stays in 
the middle branch.
In the middle branch, there are absorbers with survival amplitude 
$e^{-\xi}$ where $\xi$ assumes the value $\xi_{2\gamma}$ if the control 
photon is present and $\xi_{1\gamma}$ otherwise. 
Thus, after $N$ segments, the initial $\vec\psi_{\rm in}$ and final 
$\vec\psi_{\rm out}$ amplitudes of the target photon are related via 
\bea
\label{matrix}
\vec\psi_{\rm out}
=
\left[
\begin{array}{ccc}
\cos\epsilon & -\sin\epsilon & 0 
\\
e^{-\xi}\cos\epsilon\sin\epsilon & e^{-\xi}\cos^2\epsilon & -\sin\epsilon
\\
e^{-\xi}\sin^2\epsilon & e^{-\xi}\cos\epsilon\sin\epsilon & \cos\epsilon
\end{array}
\right]^N
\cdot
\vec\psi_{\rm in}
\,.
\ea
Now the main idea is the following: 
If the control photon is absent, absorption is very small 
$\xi_{1\gamma}\to0$ and the target photon which is initially in the 
upper branch $\vec\psi_{\rm in}=(1,0,0)^T$, ends up in the lower branch 
with high probability $\vec\psi_{\rm out}\approx(0,0,1)^T$, and vice versa.
In the presence of the control photon, on the other hand, absorption is 
much stronger, $\xi_{2\gamma}\gg\xi_{1\gamma}$, and thus the target photon 
stays in the upper or lower branch, respectively, 
due to the quantum Zeno effect. 
The above scenario describes the ideal case -- if the target photon is 
absorbed or does not end up in the designated branch, the set-up fails.
The total probability for this failure is called error probability 
$P_{\rm error}$ and should be below a given threshold. 

Since the number $N$ of segments is related to the error probability via 
$P_{\rm error}\geq\ord(1/N)$, we consider the limit $N\gg1$. 
In this limit, the desired behavior can be achieved \cite{preparation}
by setting $\epsilon=\pi/(\sqrt{2}N)$, $\xi_{1\gamma}=2P_{\rm error}/N$, 
and $\xi_{2\gamma}=\pi^2/(NP_{\rm error})$.  
As a result, the two-photon absorption rate $\xi_{2\gamma}$ 
(with control photon) and the one-photon loss rate $\xi_{1\gamma}$ 
(without control photon) have to satisfy the constraint \cite{preparation}
\bea
\label{constraint}
\kappa
:=
\frac{\xi_{2\gamma}}{\xi_{1\gamma}}
=
\frac{\pi^2}{2P_{\rm error}^2}
\gg1
\,.
\ea
For the set-up discussed in \cite{FRANSON_AND_LEUNG}, 
the corresponding error probability 
would be $P_{\rm error}=4N\xi_{1\gamma}+2\pi^2/(N\xi_{2\gamma})$ and thus 
the required value for $\kappa$ would be 64 times larger than 
in Eq.~(\ref{constraint}).
This is due to the fact that the absorbing medium is only in the middle
branch in our set-up.
In Eq.~(\ref{constraint}) we assumed that the one-photon loss rate of the 
control photon is much smaller than that of the target photon.
This could be achieved by choosing the detuning of the control photon
(see below) to be much larger than that of the target photon.
If the two loss rates were equal, for example, the necessary ratio 
$\kappa$ increases \cite{preparation} by a factor of 25
(which is still less than 64).

The above requirement (\ref{constraint}) represents a major challenge, 
since two-photon processes are typically much weaker than one-photon 
effects. 
There have been proposals to induce strong two-photon absorption 
in resonators, cavities, or fibers \cite{TWO_PHOTON_FIBER}, for example. 
Of course, these ideas could be applied to our set-up in 
Fig.~\ref{fig:3branch} as well.
However, in the following, we shall focus on free-space propagation which 
offers some advantages compared to wave-guides 
(but also has drawbacks, of course).  

\begin{figure}[h]
\begin{center}
\psfrag{e1}{$E_1$}
\psfrag{e2}{$E_2$}
\psfrag{e3}{$E_3$}
\psfrag{1s}{1s}
\psfrag{2p}{2p}
\psfrag{3s}{3s}
\psfrag{delta}{$\Delta$}
\psfrag{omega_1}{$\omega_1$}
\psfrag{omega_2}{$\omega_2$}
%
%
%
\subfigure[]{\includegraphics[width=0.4\columnwidth]{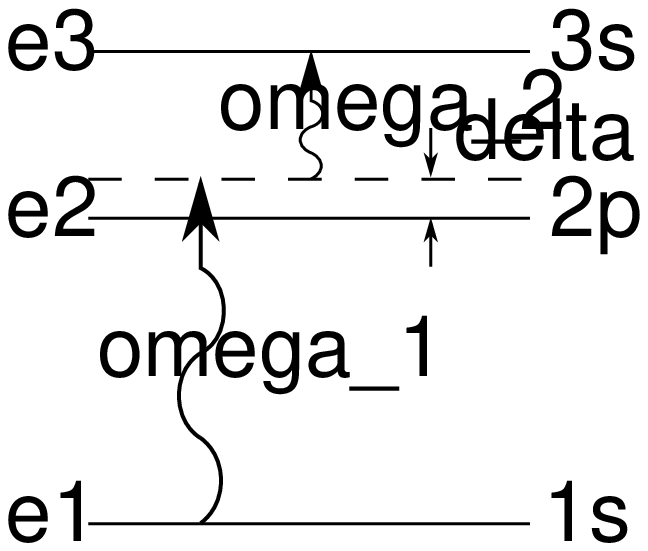}}
\hspace{.5cm}
\subfigure[]{\includegraphics[width=0.4\columnwidth]{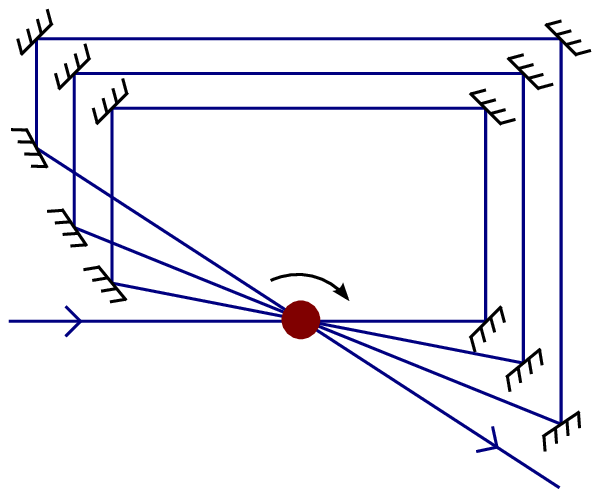}}
\caption{Sketch (not to scale) of (a) the level scheme and (b) the 
internal structure of the absorbers where the two photons are sent 
through the same absorbing medium repeatedly.
In order to ensure that the incident angle of the photons remains 
the same, the medium could be rotated.} 
\label{fig:niveau_and_repeated}
\end{center}
\end{figure}

In order to calculate (\ref{constraint}) for a concrete example, 
let us consider photons interacting with a single three-level system 
sketched in Figure~\ref{fig:niveau_and_repeated}(a). 
Starting from the usual non-relativistic Hamiltonian \cite{Scully:1997fk}
($\hbar=c=\varepsilon_0=\mu_0=1$),
\bea
\label{Hamiltonian}
\hat H
=
\frac{1}{2m}\left(
\hat{\f{p}}+q\hat{\f{A}\,}(\hat{\f{r}})
\right)^2
+V(\hat{\f{r}})
\,,
\ea
standard second-order perturbation theory \cite{Sakurai:1994uq} 
yields the two-photon absorption probability \cite{preparation}
\bea
\label{2gamma}
P_{2\gamma}
=
\frac{4\alpha^2_{\rm QED}}{\pi^2\omega_1\omega_2}\,
\frac{E^2_{12}E^2_{23}}{\Delta^2}\,
\frac{\ell_{\rm atom}^4}{A^2}
\,,
\ea
where $\alpha_{\rm QED}$ is the fine-structure constant, and 
$\omega_1,\omega_2$ are the frequencies of target and control photon, 
respectively.  
Here $E_{12}=E_2-E_1$ and $E_{23}=E_3-E_2$ are the level spacings, 
while $\Delta=E_{12}-\omega_1$ is the detuning (of the target photon). 
Finally, $A$ denotes the cross section area of the two photon wave-packets  
(we assume perfect spatial overlap), and $\ell_{\rm atom}$ stands for 
the typical dipole (coupling) length of the atomic transitions.
Note that the probability (\ref{2gamma}) does not depend on the length 
of the photon wave-packets 
(we assume perfect resonance $\omega_1+\omega_2=E_3-E_1$);
only the transversal dimensions (i.e., $A$) enter. 
In the optical regime, the contribution of the quadratic term 
$q^2\f{A}^2(\f{r})$ to  the probability (\ref{2gamma})
can be neglected.

Apart from the two-photon absorption probability (\ref{2gamma}),
we obtain scattering of single photons by the three-level system
at the same order $\ord(\alpha^2_{\rm QED})$ of perturbation theory.
In this process, the two contributions -- second order in 
$q\f{p}\cdot\f{A}(\f{r})$ and first order in $q^2\f{A}^2(\f{r})$ -- 
are of comparable strength in general. 
Under certain circumstances, such as special values of the frequencies 
$\omega_1=\omega_2=E_{12}\sqrt{1-2m\ell_{\rm atom}^2E_{12}}$, 
one could achieve destructive interference between the two contributions
leading to $P_{1\gamma}\ll P_{2\gamma}$. 
However, in the following, we shall assume that the detuning $\Delta$ 
is very small (but still much larger than the natural line width of the 
middle 2p level).
In this case, the $q^2\f{A}^2(\f{r})$ can again be neglected and the 
scattering effect from $q\f{p}\cdot\f{A}(\f{r})$ will be the dominant 
single-photon loss mechanism.
As a result, the ratio between the two-photon absorption probability 
and the one-photon scattering probability scales as \cite{preparation}
\bea
\label{ratio}
\frac{P_{2\gamma}}{P_{1\gamma}}=\frac{1}{\ord\left(\pi\omega^2_{1,2}A\right)}
\,.
\ea
In view of the diffraction limit, this ratio is smaller than one and thus 
incompatible with the requirement (\ref{constraint}).

In order to overcome this difficulty, we again exploit the quantum Zeno 
effect by sending the two photons $n$ times through the {\em same} 
absorbing medium; see Figure~\ref{fig:niveau_and_repeated}(b).
If the optical path length $L$ in between is related to the sum of the 
two photon wave-numbers via $(k_1+k_2)L\in 2\pi\mathbb N$, then the two-photon
absorption amplitudes add up coherently.
In contrast, assuming that this length $L$ is larger than the size of the 
photon wave-packets, the scattering process behaves incoherently, i.e., 
like repeated measurements, and there only the probabilities add up.  
In this way, we can enhance the total two-photon absorption probability
by a factor of $n^2$ compared to the one-photon scattering losses, 
which scale with $n$.
Furthermore, possible local one-photon absorption effects do also 
violate the phase matching requirements because 
$k_{1,2}L\not\in 2\pi\mathbb N$
and thus their probability does not scale with $n$ at all.  
In this way, the requirement (\ref{constraint}) could be achieved 
by sufficiently large $n$,
\bea
\label{ratio-n}
\kappa
=
\frac{\xi_{2\gamma}}{\xi_{1\gamma}}
=
\frac{n}{\ord\left(\pi\omega^2_{1,2}A\right)}
\,.
\ea
As an additional option, let us consider the coherent excitation 
(i.e., Dicke states \cite{Dicke:1954fk}, as often investigated regarding 
collective spontaneous emission, also known as Dicke super-radiance 
\cite{DIRECTED}) 
of a large number of atoms or molecules.
After integrating out the middle (2p) level resulting in an effective 
coupling constant $g$, cf.~Eq.~(\ref{2gamma}), the effective Hamiltonian 
of $S$ atoms or molecules at positions $\f{r}_\ell$ reads 
\bea
\label{Dicke}
\hat H
=
g\hat a_1\hat a_2\sum_{\ell=1}^S
\sigma_\ell^+\exp\{i\f{r}_\ell\cdot(\f{k}_1+\f{k}_2)\}
+{\rm H.c.}
\,,
\ea
where $\sigma_\ell^+=\ket{\rm 3s}\bra{\rm 1s}$ is the ladder operator 
for $\ell$th atom and $\hat a_1,\hat a_2$ are the annihilation operators 
of the control and target photons with wave-numbers $\f{k}_1,\f{k}_2$, 
respectively. 
In terms of the effective spin-$S$ operators 
$\Sigma^\pm=\Sigma^x\pm i\Sigma^y$ generating the $SU(2)$ algebra 
\cite{Lipkin:2002fk}, the effective Hamiltonian can be written as 
$g\hat a_1\hat a_2\Sigma^++{\rm H.c.}$
As a result, the transition matrix elements increase with the number $s$ 
of excitations, $\Sigma^+\ket{s}=\sqrt{(S-s)(s+1)}\ket{s+1}$.
Assuming $s\ll S$, the two-photon absorption probability scales with the
number $S$ of atoms or molecules times the number $s$ of excitations -- 
whereas other processes, which do not satisfy the phase matching conditions, 
only scale with $S$ and thus are suppressed for $s\gg1$. 
There is only one single-photon process \cite{preparation} which does also 
scale with $sS$, namely, the absorption of, say, the target photon with 
$\f{k}_1$ followed by the emission of a higher energetic photon with 
$2\f{k}_1+\f{k}_2$.
Fortunately, however, this process does not scale with $1/\Delta^2$ 
as in Eq.~(\ref{2gamma}), since the the ``virtual'' intermediate state 
is far from the middle (2p) level, 
and thus can be suppressed for small detuning. 

The coherent state $\ket{s}$ could be sustained by two pump lasers with 
wave-numbers $\f{k}_1',\f{k}_2'$, as long as they satisfy 
$\f{k}_1'+\f{k}_2'=\f{k}_1+\f{k}_2$ and 
$\omega_1'+\omega_2'=\omega_1+\omega_2$, in order to
fulfill the same spatial and temporal phase matching conditions as 
the control and target photon. 
%
%
Their intensities $I_1,I_2$ are related to the ratio $s/S$ via 
\cite{preparation}
\bea
\label{intensity}
\frac{s}{S} =  
\left(
\frac{4\pi\alpha_{\rm QED}E_{12}E_{23}\ell_{\rm atom}^2}
{\omega_1'\omega_2'\left(\omega_1'+\omega_2'\right)\Delta'}
\right)^2
I_1I_2
\,.
\ea
However, the detuning $\Delta'$ cannot be chosen too small in order 
to avoid unwanted excitations of the middle (2p) level,
$\Delta'\gg\sqrt{4\pi\alpha_{\rm QED}I}\ell_{\rm atom}$. 
With a typical dipole length of six Bohr radii, $\ell_{\rm atom}=6a_{\rm B}$, 
and for a large but possibly realistic intensity of 
$I = 10^{10}\,\text{W}/\text{cm}^2$, this translates into 
$\Delta'>10^{14}\,\rm Hz$. 

Let us insert some example parameters.
We assume that target and control photon are in the optical regime
(say around 500~nm) and adjust the detuning of the target photon to be 
$\Delta=3\cdot10^{12}\,\rm Hz$, which is several orders of magnitude 
larger than the typical natural line width of the middle (2p) level 
(below GHz). 
Choosing the detuning of the control photon to be an order of magnitude 
larger, i.e., $3\cdot10^{13}\,\rm Hz$, the loss rate of the control 
photon can be neglected.
For three different values of the error probability $P_{\rm error}$, 
the Table~\ref{tabelle} shows possible realizations of the set-up in 
Figure~\ref{fig:3branch} with a number $N$ of segments -- where each 
segment has a two-photon absorption probability $P_{2\gamma}^{\text{segm}}$ 
and a one-photon loss probability $P_{1\gamma}^{\text{segm}}$, which 
correspond to the ratio $\kappa$.
Assuming a focus at the diffraction limit $A = \left(\lambda/2\right)^2$, 
the last column gives the number $n$ of repetitions necessary to 
reach this ratio $\kappa$ via the mechanism sketched in 
Figure~\ref{fig:niveau_and_repeated}(b). 
Combining this enhancement mechanism with the coherent excitation of a 
large number $s$ of atoms or molecules, the number in this column then 
corresponds to $ns$. 

\begin{table}[ht]
\begin{tabular}{|c|c|c|c|c|c|c|}
\hline
$\,P_{\rm error}\,$ & $\;N\;$ & 
$\,P_{2\gamma}^{\rm segm}\,$ & $\,P_{1\gamma}^{\rm segm}\,$ & 
$\,\kappa=\xi_{2\gamma}/\xi_{1\gamma}\,$ & $\,n$ or $ns\,$ \\ 
\hline
 50\% & 10 & 95\% & 23\% & 12 & 635 \\
 25\% & 25 & 95\% & 4\%  & 75 & 3~970 \\
 10\% & 60 & 98\% & 0.5\% & 760 & 40~231 \\
\hline
\end{tabular}
\caption{Example values of the number $N$ of segments, 
the two-photon absorption probability $P_{2\gamma}^{\text{segm}}$ and 
the one-photon loss probability $P_{1\gamma}^{\text{segm}}$ per segment, 
the corresponding ratio $\kappa$, as well as the enhancement factor 
$n$ or $ns$ necessary for reaching this ratio.}
\label{tabelle}
\end{table}

Note that these are example values with some margin for trade-off.
For instance, $P_{\rm error}=50\%$ can also be reached by a ratio 
$\kappa=9$ if we increase the number of segments to $N=22$. 
This would lower the required number for $n$ or $ns$ down to 475. 
To see which values of $s$ might be realistic, let us insert a detuning 
of the pump beam of $\Delta'=3\cdot10^{14}\,\rm Hz$ into 
expression~(\ref{intensity}), where we obtain an excitation ratio 
of $s/S=2\cdot10^{-5}$. 
If we imagine a glass plate of 10~$\mu$m thickness, for example, 
a volume with a cross section area of $A = \left(\lambda/2\right)^2$ 
would contain about $2\cdot10^{10}$ molecules.
If one percent of these molecules is optically active, we get 
$S=2\cdot10^{8}$, and thus $s=4~000$. 
By inserting the aforementioned values for the detuning of the 
target photon ($\Delta=3\cdot10^{12}\,\rm Hz$) and the 
control photon ($3\cdot10^{13}\,\rm Hz$), the two-photon absorption 
probability in (\ref{2gamma}) would be around $P_{2\gamma}=\ord(10^{-11})$, 
which roughly fits to the value of $S=2\cdot10^{8}$ discussed above.

Going to the limit, we may imagine increasing these two detunings to 
$\Delta=3\cdot10^{13}\,\rm Hz$ for the target photon and 
$3\cdot10^{14}\,\rm Hz$ for the control photon  
(this value is in the infra-red region).
In this case, the two-photon absorption probability in (\ref{2gamma}) 
is two orders of magnitude lower and we could use a glass plate of 
1~mm thickness, which increases the maximum number $s$ of excited 
atoms or molecules by two orders of magnitude. 
The values in Table~\ref{tabelle} remain the same with the exception of the 
last column, where the required numbers for $n$ or $ns$ increase by 24\%. 

For the amplification mechanism sketched in 
Figure~\ref{fig:niveau_and_repeated}(b), a very tight focus is desirable.
For the other enhancement mechanism based on Eq.~(\ref{Dicke}), however, 
the spatial phase matching conditions become problematic if we focus down 
to the diffraction limit $A = \left(\lambda/2\right)^2$ due to the 
uncertainty in $\f{k}$.
Therefore, let us consider increasing the cross section area $A$. 
On the one hand, this would decrease the ratio (\ref{ratio}) even 
further -- but, on the other hand, the number $S$ of atoms or molecules
within this area $A$ grows by the same factor.
If we keep the pump laser intensity constant, the enhancement factor 
$s$ compensates the shrinking ratio (\ref{ratio}), i.e., a weaker 
focus with smaller uncertainty in $\f{k}$ is also feasible.

In order to explore another direction in the multi-dimensional
parameter space, one could consider replacing the atoms or molecules
with artificial atoms such as quantum dots, which generate electronic 
bound states with a level structure similar to 
Figure~\ref{fig:niveau_and_repeated}(a). 
In this case, the dipole length $\ell_{\rm atom}$ could be increased 
by several orders of magnitude, see Eq.~(\ref{intensity}), but the
realistic energy and intensity scales would also have to be modified. 

In summary, we have presented a scheme for entangling photons via double 
application of the quantum Zeno effect, which might be realizable with 
present day technology.
Note that we consider the (pseudo) deterministic creation of 
entanglement between two {\em given} photons, which is very different 
from the {\em spontaneous} creation of entangled photon pairs via 
parametric down conversion.
Our scheme is based on free-space linear optics in connection with 
a two-photon absorbing medium.
Even though the example parameters above show that it might be hard 
to directly reach the error threshold of $10^{-4}$ or $10^{-5}$
required for universal quantum computation, the achievable success 
probabilities for entangling photons are already comparably high.
In addition, by placing photon detectors around the apparatus, 
one can detect occurring errors with high probability and thus 
apply error correcting schemes; see, e.g., \cite{MYERS_AND_LEUNG}.


$^*$ {\tt ralf.schuetzhold@uni-due.de}



\end{document}